*Oh sister, where art thou? Spatial population structure and the evolution of an altruistic defence trait*

T. Pamminger[1,2], S. Foitzik[1], D. Metzler[3] and P. S. Pennings[4]

[1]Institute of Zoology, Johannes Gutenberg University of Mainz, Germany

[2]School of Life Sciences, University of Sussex, Brighton, UK

[3]Department of Biology II, Ludwig Maximilian University of Munich, Planegg – Martinsried, Germany

[4]Department of Biology, San Francisco State University, San Francisco, CA, USA

**Running title:** Fitness benefits of slave rebellion

**Correspondence:** Tobias Pamminger, *School of Life Sciences, JMS building, University of Sussex, BN1 9RH, Brighton, UK*

e-mail: t.pamminger@sussex.ac.uk

# Abstract

The evolution of parasite virulence and host defences is affected by population structure. This effect has been confirmed in studies focusing on large spatial scales, whereas the importance of local structure is not well understood. Slavemaking ants are social parasites that exploit workers of another species to rear their offspring. Enslaved workers of the host species *Temnothorax longispinosus* have been found to exhibit an effective post-enslavement defence behaviour: enslaved workers were observed killing a large proportion of the parasites' offspring. Since enslaved workers do not reproduce, they gain no direct fitness benefit from this "rebellion" behaviour. However, there may be an indirect benefit: neighbouring host nests that are related to "rebel" nests can benefit from a reduced raiding pressure, as a result of the reduction in parasite nest size due to the enslaved workers' killing behaviour. We use a simple mathematical model to examine whether the small-scale population structure of the host species could explain the evolution of this potentially altruistic defence trait against slavemaking ants. We find that this is the case if enslaved host workers are related to nearby host nests. In a population genetic study we confirm that enslaved workers are, indeed, more closely related to host nests within the raiding range of their resident slavemaker nest, than to host nests outside the raiding range. This small-scale population structure seems to be a result of polydomy (e.g. the occupation of several nests in close proximity by a single colony) and could have enabled the evolution of "rebellion" by kin selection.

# Introduction

Organisms that exploit other organisms as a resource are an evolutionary success story. Parasitism is one of the most common lifestyles on earth, shaping food webs, influencing biodiversity and driving the evolution of major evolutionary innovations such as sex (Hamilton et al., 1990; Daszak et al., 2000; Wommack and Colwell, 2000; Morran et al., 2011). The high costs associated with being parasitised exert strong selection pressure on hosts to defend themselves against this exploitation (Booth et al., 1993). Indeed, host species developed various defence traits limiting the harmful effects of parasites (Minchella, 1985; Brunham et al., 1993; Gross, 1993; Bennett et al., 1994; Sheldon and Verhulst, 1996; Morran et al., 2011). Studying such responses can help to understand the factors shaping the evolution of parasite virulence and host resistance.

The dynamics of host-parasite interactions have been the focus of numerous theoretical studies (Bremermann and Pickering, 1983; Boots and Haraguchi, 1999; Gandon and Michalakis, 2002). Recently, spatial population structure has been found to affect the evolution of parasite virulence (Boots et al., 2004) and host resistance (Best et al., 2011). In fact, most populations exhibit spatial structure at least on some scale, which can influence the outcome of host-parasite interactions (Thompson, 1994; Lively, 1999; Gandon and Michalakis, 2002).

In contrast to micro-parasites, which often directly extract resources from their hosts, brood parasites exploit the hosts' brood care behaviour (Petrie and Møller, 1991). Many organisms devote a large proportion of their adult life to the energetically and time costly activity of rearing offspring. Brood parasitism is a strategy to reduce these costs, by delegating brood care to individuals other than the parents. This behaviour has evolved independently in birds, fish and insects (Sato, 1986; Rothstein, 1990; Kilner and Langmore, 2011). Slavery or dulosis is a common variant of brood or social parasitism in ants (Buschinger, 2009). Obligate slavemakers have lost the ability to perform basic tasks such as brood care, foraging and nest relocation, which are outsourced to enslaved workers of other species (Stuart and Alloway, 1985). These workers are acquired during destructive slave raids in which slavemakers steal brood from surrounding host colonies (Alloway, 1979; Foitzik et al., 2001). The brood is transported back to the parasite nest where the enslaved host workers eclose and start caring for the parasites' offspring. Ant host species have developed a wide range of defence mechanisms including enemy recognition, and fight and flight strategies (Alloway, 1979; Foitzik et al., 2001; Bauer et al. 2009; Ruano et al. 1999). These behaviours can increase the host's fitness directly by averting or reducing the impact of raids and thereby preventing or diminishing the enslavement of host workers. The focal species of this study, the ant *Temnothorax longispinosus,* is the main host of the common slavemaker *Protomognathus americanus,* an obligate brood parasite. Parasitised populations of this host are characterized by various constitutive and induced behavioural defence traits (Alloway 1990; Foitzik et al., 2001; Brandt and Foitzik 2004 Pamminger et al., 2011, Pamminger et al. 2012, Scharf et al., 2011; Kleeberg et al. 2014).

In contrast to the well-studied defence mechanisms that reduce worker enslavement, it has been argued that post-enslavement defence mechanisms cannot evolve in slavemaker systems, since the enslaved workers lack options to increase their fitness directly. For example, enslaved workers cannot simply return to their mother colony because they do not know where it is located, as they were stolen as larvae or pupae. In addition, their mother colony may not have survived the slave raid. Staged laboratory raids show that host nests lost a high proportion of their brood and workers, and often their queen (Foitzik et al., 2001). Therefore not surprisingly, genetic field studies revealed that most host nests do not survive raids under natural conditions (Foitzik and Herbers, 2001a, Blatrix and Herbers 2003). Even in the unlikely event of returning to their mother colony, enslaved workers would likely be rejected by their own colony, because they have acquired the slavemakers' chemical scent (Gladstone, 1981; Brandt et al., 2005). If enslaved workers cannot return home, they could potentially increase their fitness directly by reproducing. Like most ant workers, *Temnothorax* workers have the ability to produce male offspring (Heinze et al., 1997) and could therefore increase their fitness directly. However, enslaved *T. longispinosus* workers have never been observed to reproduce in the presence of *P. americanus* workers or queens (Foitzik and Herbers, 2001a; Foitzik and Herbers, 2001b), which apparently effectively suppress slave reproduction.

Using similar reasoning, Gladstone (Gladstone, 1981) argued that post-enslavement defence traits could not be selected for, trapping enslaved workers in an evolutionary dead end. However, Gladstone neglected the possibility that enslaved workers may be able to increase the fitness of their free-living relatives in nearby host colonies by reducing local raiding pressure. If enslaved workers could decelerate or stop the growth of slavemaking colonies, they would reduce the number and severity of slave raids on nearby host colonies (Foitzik and Herbers, 2001b). Such a post-enslavement behavioural defence was indeed observed in the ant *T. longispinosus* (Achenbach and Foitzik, 2009), who termed this behaviour "slave rebellion": Instead of rearing the slavemakers' offspring to adulthood, enslaved workers stop caring for parasite offspring once they developed into pupae. Instead, they attack the slavemaker pupae, tearing them apart or carrying them out of the nest, where they die from neglect. Slave rebellion occurs to various degrees in all host populations investigated (Pamminger *et al.*, 2013). This suggests that slave rebellion is either an ancestral defence trait or has evolved repeatedly in various populations.

In addition to the direct behavioural observation of enslaved workers killing slavemaker offspring in the laboratory, field collections of *P. americanus* colonies show that enslaved workers are indeed less productive in terms of per-capita production of new queens and workers than free-living workers in unparasitised nests (Foitzik and Herbers, 2001a). The destruction of slavemaker offspring by enslaved workers could explain the extremely small colony size of *P. americanus* colonies, of around five slavemaker workers (Foitzik et al. 2009), which is much smaller than colonies of related slavemaker species, such as *T. duloticus* or *T. pilagens* (Alloway, 1979; Seifert et al., 2014).

An alternative non-adaptive explanation for the low survival rate of *P. americanus* slavemaker brood could be that enslaved host workers are less able to care for heterospecific brood. If so, we would expect that during the time of the most intensive care – the larval stage – the highest fraction of the slavemaker brood perishes. In contrast, the reverse is the case; slavemaker larvae survive very well until pupation, when the killing of apparently healthy pupae starts (Achenbach et al. 2009). Moreover, the difference in survival rates between larvae and pupae can be explained by the availability of recognition cues, as larvae do not exhibit a distinct chemical profile, while *P. americanus* slavemaker pupae can be recognized by their species-specific cuticular hydrocarbons (Achenbach and Foitzik 2010). Finally, during coevolution slavemakers should be selected to be able to develop well under slave worker care, and a phylogeny showed that *P. americanus* is a relatively old social parasite (much older than its closest slavemaker relatives *T. duloticus* or *T. pilagens,* from which slave rebellion has not been reported) (Beibl et al. 2005). Hence all the evidence indicates that the rebellion behaviour of enslaved *T. longispinosus* workers is an adaptive trait.

In this study we develop a simple mathematical model to determine whether a hypothetical rebellion allele could establish itself in a host population. We find that it could spread if the host population is sufficiently spatially structured on a local scale, and the parasites' raiding range is limited to this local sphere. Next we use a population genetic approach to determine whether the host population is sufficiently structured within the parasites' raiding range. It is unclear, a priori, how structured the host population is on a small scale, because there are at least two processes expected to erode small-scale structure in the study system: first, the queens of *T. longispinosus* participate in mating flights (Howard and Kennedy, 2007) facilitating queen dispersal and second, *T. longispinosus* colonies inhabit ephemeral nest sites (e.g. acorns) which force them to relocate to new nest sites regularly (Herbers and Tucker, 1986). On the other hand, there is at also a reason to expect some small-scale structure: the host species is known to be at least somewhat polydomous, meaning that a single colony can occupy multiple nest sites simultaneously. Such additional nests are usually established in spring and remain connected to the main nest, both by worker and brood relocation, at least for some time (Herbers and Tucker, 1986). In addition, *T longispinosus* nests can have multiple queens (polygyny) (Herbers 1984) but it is not known exactly how such queens are related to each other. It is not clear how and to which extend polygyny influences the small-scale population structure.

To quantify the hosts' population structure on the relevant spatial scale, we first need an estimate for the slavemakers' raiding range, which we obtain by using published data and under the assumption that slavemaker preferentially raid nearby nests. This assumption is reasonable because the slavemaking ants are small (approximately 3 mm) and lack sophisticated trail pheromones to organize long distance recruitment (Alloway, 1979). To investigate potential indirect benefits of the slave-rebellion behaviour we mapped the small scale population structure in two well-studied communities (Foitzik et al., 2001; Foitzik et al., 2009) in the Eastern United States (New York and West Virginia) and genotyped *T. longispinosus* host workers in nests with and without slavemakers to estimate relatedness between enslaved workers and surrounding host colonies. The aims of this study are:

1. to determine whether under relevant assumptions, a hypothetical slave rebellion allele could invade a population of non-rebels, by developing a simple mathematical model.
2. to test whether and to which extent enslaved workers are related to surrounding host colonies within the raiding range of slavemaker nests.
3. to determine whether slave-rebellion behaviour could lead to indirect fitness benefits.
4. to quantify the small-scale population structure of the host species, in order to investigate why enslaved workers could be related to surrounding host colonies.

**Methods**

**The model**

We developed a simple mathematical model to analyse whether a hypothetical "slave rebellion" allele could invade a population of non-rebels (or, in other words, the absence of slave rebellion is evolutionarily unstable), and to infer which parameters may play a crucial role in the evolution of this trait. For this purpose, we assume that a hypothetical dominant rebel allele is so rare that we can ignore any small probability that there may be neighbouring rebel nests when we calculate the fitness of non-rebels.

The main purpose of our model is to render explicit our hypotheses about the principal mechanism through which rebellion may be beneficial. Making accurate quantitative predictions with a more detailed model would be possible with a vast amount of data, which is currently not available. We rather aim for a simple model to clarify the impact of the most crucial parameters on a qualitative level. To keep our model as transparent as possible, we neglect the impact of nest size, age, and of the number of rebels involved in a rebellion. Also, for simplicity, we ignore the effect of queen presence or absence in a nest, assuming that each nest has the same expected lifetime productivity

(see discussion for some remarks on this assumption). Furthermore, we make the following assumptions:

1. Host and slavemaker nests occur in neighbourhoods. The slavemakers' raiding range determines the size of the neighbourhood.

2. All host nests in a neighbourhood have an equal probability, $p$, to be raided in a raiding season, if a slavemaker nest is present in the neighbourhood.

3. Neighbourhoods of different slavemaker nests do not overlap.

4. If a host nest is raided in a certain year, it will survive all raids of this year with probability $z$.

5. If a slavemaker nest raids a nest containing rebels, the slavemaker nest will die with probability $r$ at the end of the year. If it survives the year, it will begin the next year as a rebel-free slavemaker nest.

6. Host nests are founded in slavemaker-free neighbourhoods.

7. A slavemaker-free neighbourhood has a fixed probability, $s$, to be inhabited by a slavemaker nest in the next season.

8. Host nests containing rebels have an average of $n$ related nests in the neighbourhood that also contain rebels. (we do not account for the precise fraction of rebels among the workers.) The parameter $n$ summarises the effects of two other parameters: the average size of clusters of related nests and the relatedness between these nests.

9. For computing the productivity of host nests that do not carry the rebel allele we can neglect the fraction of nests that have rebel nests in their neighbourhood, as the rebel allele is assumed to be rare.

10. There is a cost, $c$, associated with carrying the rebel allele.

11. Host nests and slavemaker nests can die of natural causes (not due to a raid or rebellion) with probability $(1-u)$ and $d$ respectively. If the slavemaker nest dies and is immediately replaced by another slavemaker nest, we count that as if the first slavemaker nest would have survived. Thus, *d* (and in case of rebellion *r*) is in essence the probability that the slavemaker nest dies and is not replaced by another slavemaker nest.

An overview of the model parameters is given in Table 1. We calculated the expected lifetime of host nests with and without rebels by first step analysis (Pinsky and Karlin, 2011). Since we neglect size and age of nests, we use lifespan as a proxy for productivity of a nest and thus direct fitness. In our setting, let $t_1$ be the expected time (in years) until a host nest dies in a slavemaker-free neighbourhood, and let $t_2$ be the expected time if there is a slavemaker nest in the neighbourhood. A host nest in a slavemaker-free neighbourhood will live to the next season with probability *u*. Then, if with probability 1-*s* no slavemaker arrives, the expected residual lifetime of the host nest from the start of the next season will still be $t_1$. If a slavemaker nest arrives during the season, the host nest's residual lifetime after the first year will be $t_2$. Thus, we obtain the equation $t_1 = u \cdot (1 + (1-s) \cdot t_1 + s \cdot t_2)$. For $t_2$, we account for the possibility that the host nest can die due to a raid, and that for the next year, the slavemaker nest can die or survive (or die and be replaced by another slavemaker nest). We obtain $t_2 = u \cdot (1 - p \cdot (1-z)) \cdot (1 + d \cdot t_1 + (1-d) \cdot t_2)$. Taking all model parameters as fixed values, we now have two equations for the two unknowns $t_1$ and $t_2$. Solving this linear equation system is straightforward, but the general solution looks complex and is not instructive, hence it is not shown. Now, let $t_3$ and $t_4$ be the expected times until a rebel nest dies if slavemakers are absent or present, respectively, in the neighbourhood. The equation for $t_3$ is analogue to that for $t_1$: $t_3 = u \cdot (1 + (1-s) \cdot t_3 + s \cdot t_4)$. In the equation for $t_4$ we account for the possibility that the slavemaker has an increased death risk if it has raided either the focal rebel nest (which happens with probability *p*) or one or more of the other *n* rebel nests in the neighbourhood (which has probability $1-(1-p)^n$ ). We obtain:

$$t_4 = u \cdot [1 - p \cdot (1-z) + p \cdot z \cdot (r \cdot t_3 + (1-r) \cdot t_4) + $$
$$(1-p) \cdot ([(1-p)^n \cdot d + (1-(1-p)^n) \cdot r] \cdot t_3 + [(1-p)^n \cdot (1-d) + (1-(1-p)^n) \cdot (1-r)] \cdot t_4)]$$

Again, we can solve the linear equation system with two equations to obtain $t_3$ and $t_4$. As we assume that host nests are usually founded at sites without slavemakers, then $t_1$ is also the expected total lifetime of a non-rebel host nest, and $t_3$ is the expected lifetime of a rebel nest. We assume that the rebel allele can spread in the population if $(1-c) t_3 > t_1$ or, equivalently, $K := (1-c) \cdot t_3 / t_1 > 1$. In other words: the longer lifespan of rebel nests as compared to non-rebel nests must at least compensate for the reduced productivity due to the cost of the rebel allele. We calculated the influence of all model parameters for rebel allele costs of 0.05, 0.1 and 0.2. When we varied one parameter, we set the other parameters to the default values defined in Table 1.

**Ant collection and genotyping**

*Temnothorax longispinosus* host and *P. americanus* slavemaker colonies were collected in July 2009 at the Huyck Preserve in Albany County, New York, USA (N42°31'35.3" W74°9'30.1") and at the Watoga State Park, Pocahontas County, West Virginia, USA (N38°06'13" W80°8'59"). To investigate small-scale genetic structure, we mapped two transects per habitat consisting of six plots (6×3 m) each, spread over a distance of about 100m (Fig.1). In West Virginia (WV) we mapped two additional plots separated from the two transects. We carefully searched the leaf litter in all plots and inspected all potential nest sites such as rotten twigs and acorns. We recorded nest locations within each plot and distances between plots, which enabled us to calculate distances between all nests in a given transect. Ant nests were censused directly after collection. We recorded queen number, number of workers and number of brood including larvae and pupae (but not eggs). Ant colonies were transferred into 100% ethanol for later genetic analysis. We amplified six highly variable microsatellite loci (for DNA extraction and PCR protocols see electronic supplementary material) for five workers per non-parasitised *T. longispinosus* nest and 20 enslaved *T. longispinosus* workers per *P. americanus* nest (or as many as present in case fewer workers were found in the colony). We chose a larger sample size of workers in slavemaker nests relative to workers from non-parasitised nests, because enslaved workers in slavemaker nests typically originate from several host nests, and are therefore expected to be much more genetically diverse (Foitzik et al., 2001). In total we collected genetic data from 1181 non-parasitised *T. longispinosus* workers (from 241 *T. longispinosus* nests, on average 4.9 individuals per nest) and 683 enslaved *T. longispinosus* workers (from 41 *P. americanus* nests, on average 16.7 individuals per nest). We will refer to *T. longispinosus* workers from non-parasitised nests as "free-living" and to the enslaved *T. longispinosus* workers sampled from *P. americanus* nests as "enslaved".

**Kin-selected benefits of rebellion: Relatedness of slaves to surrounding hosts**

**Raiding distance**

It is hard to measure raiding distance in the field by observing slave raids of *P. americanus*, because a) the ants are small (~3mm), b) only few ants participate in raids (~10 ants) and c) the raids take place in dense leaf litter on the forest floor. We therefore used a method based on published raiding frequencies and density data. To estimate the approximate area threatened by raids of a slavemaker nest we divided the average number of successful raids a slavemaker nest conducts each season by the average density of *T. longispinosus* colonies in parasitised areas (Foitzik and Herbers, 2001a; Foitzik et al., 2001, 2009). This gives us the minimum area raided by a slavemaker nest each season because, for example, to conduct five (NY) successful raids per season a slavemaker nest has to cover at least an area containing five colonies. Assuming no directional preference of the slavemaker we can calculate the approximate raiding range around a slavemaker nest using the radius $r = \sqrt{(A/\pi)}$ of a circle of area A which is the area threatened by the slavemakers. We assume that host and slavemaker nests stay in one location during the raiding season and that other host nests do not recolonize empty nest sites during this time.

**Relatedness of enslaved workers to nearby free nests**

We cannot determine whether enslaved workers and nearby host nests share a rebellion allele, because we have not identified such an allele. To estimate the probability that they share a (rare) rebellion allele, we used relatedness values estimated from neutral genetic markers (microsatellites), in order to determine whether enslaved workers are more related to nearby host nests than host nests further away.

After estimating the raiding range, we analysed whether host nests within the slavemakers' raiding range are more closely related to the enslaved workers in this nest compared to other nests in the same transect. This analysis is important because it determines whether a slave's behaviour can impact its indirect fitness by changing the raiding risk for related colonies. To test whether nearby nests are related to enslaved workers, we computed the coefficient of relatedness of enslaved workers to surrounding free nests in the same transect. The test statistic is the average of these relatedness values over all cases in which the free nest is within the raiding range of the slavemaker nest in which the enslaved workers were found. Relatedness was computed in R (R Core Team, 2012) according to equation (10) in Queller-Goodnight (1989). Our R script is available upon request. We performed a permutation test similar to a Mantel test with 1000 repetitions (Mantel, 1967). In each repetition we shuffled the nests within each transect of a sampling area (NY or WV) for the computation of distances and computed the test statistic using these permuted distances and the original coefficients of relatedness. The one-tailed p-value is the relative rank of the original value of the test statistic within those from the 1000 nest permutations.

**Small scale population structure of the host species**

**Relatedness between free-living host nests**

In a next step, we investigated why enslaved host workers are more closely related to nests within the raiding range of their slavemaker nest. To find the origin of the observed increase in relatedness we had a closer look at the free-living host populations. To see whether we could detect a similar small-scale structure we calculated the pairwise relatedness between free-living host nests and tested for small scale population structure, utilizing the same test statistic as in the previous section. In addition, we also performed a classical Mantel test (Mantel, 1967) as in Trontti et al. (Trontti et al., 2005a), but based on Jost D (Jost, 2008) values (as opposed to coefficients of identity).

**Source of host population structure: polydomy**

A potential source for small-scale population structure in *T. longispinosus* population could be polydomy, i.e., the occupation of several nest sites by a single colony. To identify polydomous nest pairs we assumed that polydomy arises when a random subset of individuals in a nest move to a new nest site. The relatedness between polydomous nest parts can then be estimated by splitting the five sampled individuals from free-living nests into a group of two and a group of three individuals, and calculating the relatedness between those two groups according to Equation (10) in Queller and Goodnight (Queller and Goodnight, 1989). This was done for all free-living colonies for which we had five sampled individuals. We next fitted a beta distribution to those relatedness values, whose domain was stretched to the interval between -1 and 1 by linear transformation. We will refer to this distribution as the "distribution $D_{within}$ of relatedness within nests" (Fig. 2 a, b). Beta distributions are the most commonly used class of distributions defined on a continuous finite interval. The fitted distribution of $D_{within}$ has a mode of 0.55 in NY and 0.65 in WV. To determine relatedness between nests that are not part of a polydomous colony, we took pairs of nests that were more than 50 meters apart and calculated relatedness between two random individuals from one nest and three from another nest. We used two and three individuals to maintain consistency with the same sample sizes as before, again using the same algorithm (Queller and Goodnight, 1989). We then fitted a second beta distribution, which we will refer to as "distribution $D_{indep}$ of relatedness between independent nests" (Fig. 2 c, d), which has a mode of close to 0 in both study sites.

Next we divided all nest pairs into six distance segments: <1m, 1-2m, 2-3m, 3-5m, 5-10m and >10m. For each segment we modelled the distribution of relatedness values found within this segment as a mixture between the distributions $D_{within}$ and $D_{indep}$. This means that for each pair of nests, we assumed that with probability p its relatedness comes from the distribution $D_{within,}$ and with probability 1-p it comes from the distribution $D_{indep}$. The mixture coefficient p is fitted to the relatedness data of nests in the distance segment (see Fig. 2 e, f). We interpret p as the probability that two nests, found with such a distance, belong to the same polydomous colony.

For each of the distance bins, the weights of the two distributions define a threshold, the relatedness value where the distribution $D_{within}$ of relatedness within nests rises above the distribution $D_{indep}$ of relatedness between independent nests (indicated by the blue dotted line in Fig. 2 e, f). Nest pairs within the distance bin with a relatedness value higher than this threshold are likely to belong to the same polydomous colony (see Fig. 3 for thresholds). At more than 2 meters distance, most colony pairs are not polydomous, but at closer distances there are a number of colonies which are likely polydomous (indicated as solid dots in Fig. 3).

Next, for each host nest $x$ we estimated the number $n_x$ of nests that (1) had a distance of less than 1 m from $x$, (2) were genotyped and (3) belonged to the same polydomous colony as $x$. By summing up for all nests that fulfil (1) and (2) the probability that (3) is also fulfilled, as estimated by fitting the beta distribution model. The average value of $n_x$ (averaged over all genotyped nests $x$) then serves as conservative estimate of the parameter $n$ in our theoretical model (Table 1). This estimator is conservative for three reasons. First, nests of the same polydomous colony could have a distance larger than 1 m, and it appears possible that host nests with a distance larger than 1 m belong to the same neighbourhood raided by a slavemaker nest. Second, for nests $x$ that are closer than 1 m to the boundary of a plot, we have not genotyped all nests $y$ closer to $x$ than 1 m. Third, when fitting the beta distribution model, we assume that the relatedness of nests of the same colony is as high as the within-nest relatedness, *e.g.* because they stem from the same queen or stem from related queens and exchange workers. In this case it is highly probable that if the workers of one nest have a rebel allele, the same holds for the workers of the other nests of the same colony. It may be possible, however, that neighbouring nests which do not belong to the same colony in our strict sense, are still more closely related than to a distant nest. For example, this is the case if one nest is founded by a queen stemming from the other nest and there is no exchange of workers between the nests. In this case there could still be an increased probability of sharing a rare allele.

## Results

### Model predictions

Our model predicts that the fitness of the hypothetical rebel allele (relative fitness $K$ compared to non-rebels), increases with the probability that a slavemaker nest appears in the proximity of a host nest (Fig.4). Even for low costs of 0.05, the rebel allele could not spread if the probability that a slavemaker nest arises is close to 0 (Fig. 4a) or if slavemaker nests have high probabilities to die even without rebellion (Fig. 4b). The chance of a rebel allele to spread in a population is highest if the probability $z$ of host nests to survive raids is intermediate (Fig. 4c). If it is too low, the free-living host nests are all killed before they can benefit from rebellion, and if $z$ is too high, the benefit of the rebel allele is small. Moreover, the fitness of the rebel allele depends on the efficacy of rebellion, that is, the probability $r$ that it leads to the death of the slavemaker nest (Fig. 4d). Intermediate values of $p$ (the likelihood of a host nest to be raided in the neighbourhood of a slavemaker nest) also led to the highest chances of the rebellion allele to spread (Fig. 4e). The reasons may be similar as for parameter $z$. Furthermore, if rebel nests tend to be surrounded by at least a few other rebel nests, it clearly enhances the fitness of a rebel allele, even though the fitness quickly reaches a plateau with increasing $n$ (Fig. 4f). Finally, the advantage of rebellion increases with the general probability $u$ of host nests to survive a year, as this increases the chances that host nests have a long-term advantage of the death of the nearby slavemaker nest (Fig. 4g).

Our model suggests that a slave rebellion allele could invade a population via indirect fitness benefits, if host colonies are structured in clusters of related nests within the raiding range of the slavemaker nest. We calculated the approximate raiding range of slavemaker nests using data from earlier studies, which show that slavemaker colonies conduct a minimum of five (NY) and two (WV) successful slave raids per season on host colonies (Foitzik and Herbers, 2001b; Foitzik et al., 2001). In addition, we know that the average host nest density in the two investigated habitats is 0.90 (NY) and 0.59 (WV) host nests/m$^2$ (Foitzik et al., 2009). Using this data we estimated the area threatened by a single slavemaker colony to be a minimum 5.5 m$^2$ in NY and 3.4 m$^2$ in WV, which corresponds to a raiding range with a radius of 1.32 m (NY) and 1.04 m (WV) around a slavemaker nest. Host nests within this radius of a slavemaker nest have a very high probability to be raided in a given season.

**Relatedness of enslaved workers to nearby free host nests**

The average relatedness (r) of enslaved workers to free-living host nests found within the raiding range of the slavemaker nests of 1.32 m in NY and 1.04 m in WV was r = 0.07 in New York r = 0.14 in West Virginia (Fig. 5, difference between NY and WV not significant, p = 0.3). Both values are higher than the relatedness of enslaved workers to host nests found outside the raiding range in the same transect (mean relatedness outside the raiding range r = 0.021 in NY and r = -0.01 in WV, permutation test p = 0.013 in NY and p = 0.001 in WV, Fig. 5). The coefficients of relatedness are clearly increased in the close vicinity of slavemaker nests, which means that enslaved workers and nearby nests may also share a hypothetical rebellion allele and that there is opportunity for enslaved workers to increase the fitness of related nests through rebellion.

**Small scale population structure of the host species**

**Relatedness between free-living host nests**

We analysed whether the free-living host population exhibited elevated relatedness values between close neighbours, similarly to enslaved workers and their free-living neighbours. We find small-scale structure in both habitats, consistent with our model assumptions. Closely neighbouring nests (i.e. nests located within a distance smaller than our estimated raiding range) are more closely related to each other than nests that are located further apart, both in New York (p<0.001) and in West Virginia (p<0.001; Fig. 3). In NY, relatedness values between neighbouring free-living nests resemble those between enslaved workers and free-living neighbours (mean relatedness within raiding range: r = 0.08 in free-living colonies versus r = 0.07 for enslaved workers-free comparison). In WV, relatedness values between neighbouring free-living nests are tentatively higher than those between enslaved workers and free neighbours (mean relatedness within raiding range: r = 0.22 in free-living colonies versus r = 0.14 for enslaved-free comparison in WV, Fig. 3 and 5). However, in both cases the difference is not significant (both p > 0.19). Thus, local structure amongst free-living nests is sufficient to explain why enslaved workers are more closely related to nearby host nests. Based on the beta distribution model of relatedness values (Fig. 2) we obtained conservative estimates for the model parameter *n* (Table 1), the average number of additional nests occupied by a rebel colony. For the data sampled in WV, this estimate was 0.56, and for the data sampled in NY it was 0.36. The average probability of belonging to the same colony for nest pairs as predicted by the model decreased with distance and was higher in West Virginia (< 1 m: 0.427, 1 - 2 m: 0.072, 2 - 3 m: 0.036) than in New York (< 1 m: 0.147, 1 - 2 m: 0.071, 2 - 3 m: 0.008).

A classical isolation-by-distance analyses (Mantel test (Mantel, 1967) using Jost's D (Jost, 2008)), provides evidence for isolation-by-distance in NY, but not in WV (New York: Pearson correlation for geographic vs. genetic distance r = 0.06, P<0.001; West Virginia: r = 0.01 P=0.54, methods as in Pennings *et al.*, 2011).

**Source of host population structure**

To see whether the observed small-scale population structure is due to polydomy, we eliminated the nest pairs which we identified as probably polydomous from the data set, and determined the average relatedness between the remaining nests at distances below the raiding range. After eliminating polydomous nests we no longer detect a significant population structure (NY mean relatedness near: 0.04, far: 0.02, p=0.6 WV near: 0.004, far: 0.01, p=0.8). Our results thus indicate that polydomy is sufficient to explain the small-scale population structure in host populations.

**Discussion**

In *On the Origin of species*, Charles Darwin referred to the evolution of sterile social insect worker traits as one "special difficulty" that appeared fatal to his entire theory of natural selection. Indeed, the question of how altruistic behaviours in social insects arose is still under debate (Nowak et al. 2010, Abbot et al. 2011). "Slave rebellion" ties into this discussion as it presents a special case of this difficulty, since enslaved ant workers do not reproduce, and moreover, they do not even interact directly with their relatives. Here we analysed whether the rebellion behaviour of enslaved ant workers, which leave no direct offspring, could be selected for by kin selection if surrounding host colonies would profit from this behaviour and are related to the slaves. It is known that slavemaker nest size is positively correlated with raiding frequency and destructiveness (Foitzik and Herbers 2001b). Therefore, by reducing the number of slavemaker workers in a nest, enslaved workers could lower the raiding pressure on surrounding host nests. Based on a simple mathematical analysis, we show that slave rebellion can result in an indirect fitness benefit, if the costs of rebellion are low and enslaved workers are related to nearby host nests. Using spatial and population genetic data we then confirm that enslaved workers are related to nearby host nests in both studied populations (New York and West Virginia) and that polydomy (the occupation of multiple nest sites by one colony) is sufficient to explain the local structure that we observe.

Our theoretical analyses provide more detailed insights into the parameter space under which slave rebellion could evolve. First, we show that the rebellion behaviour only provides a benefit if there is a substantial probability that a host nest is raided before it dies of natural causes. This is the case when there is a high probability *s* for a slavemaker to move into the area, and when the death rate ($1-u$) of host nests that are not affected by slavemakers is low. Clearly, these two conditions increase the host nests' benefit of being protected against slavemakers.

Second, for the rebel allele to have an effect, the cluster size (number of nest sites used by a colony) needs to be larger than 1 (which in the model means n>0). This is required since only if one nest in a cluster of rebel nests is raided and at least one nest is not raided, the surviving host nest(s) can benefit from the rebel behaviour of the enslaved workers if the slavemaker colony will conduct fewer raids the next year. According to our model predictions, the presence of one related nest in a neighbourhood brings about a large benefit for the rebel allele, but larger numbers of related neighbouring nests only slightly increase this benefit (Fig. 4f). We estimated $n$ (the average number of related nests) from the genetic data sampled in NY and WV and obtained values of 0.36 and 0.56, respectively. For these values our model predicts that slave rebellion behaviour could spread in the host population if the cost of rebellion is around 10% or lower, depending on the values of the other parameters. However, it must be noted that as our model is designed to be simple and transparent, we do not expect that its predictions are quantitatively accurate.

Finally, since slave rebellion only has an effect if at least one nest of a cluster is raided and at least one nest survives, intermediate raiding probabilities ($p$) lead to the highest fitness benefits (Fig. 4e). If $p$ is too high, it is probable that all nests are raided, with no survivors that could benefit from the rebellion behaviour. If $p$ is too low, it is probable that no nests are raided, so that there are no enslaved ants that could rebel. In addition, for the probability $z$ that a host nest survives a raid, it is also the intermediate range of values that makes slave rebellion most effective (Fig. 4c). Rebellion can only be of direct benefit for the rebel's home nest if there is a chance that the nest has survived the raid. This is why for small values of $z$, an increase of $z$ may increase the benefit of slave rebellion. Yet if $z$ is close to 1, the cost of being raided is small, which reduces the benefit of raid reduction by slave rebellion.

Note that we have assumed that all host nests that make up a cluster of nests have the same expected productivity. In reality this may not be the case. Specifically, one nest may be the main nest with a queen, whereas another nest may be queenless. If the queenright nest is raided and doesn't survive, productivity of the other nests in the cluster may suffer dramatically. In this case the benefit of rebellion would be smaller than predicted by our model. On the other hand, if the queenless nest is raided, the productivity of the other nest may not be much affected. This case could be more beneficial for the rebel allele than what we consider in our model. Therefore the fact that there are queenright and queenless nests in a cluster seems to generate opposing outcomes, such that these two effects may cancel each other out.

Using small scale population genetic data, we found that both in NY and in WV, enslaved workers are more closely related to nearby free-living ants than to free living ants located further away (Fig. 5). This result indicates that enslaved workers and nearby nests may also share a hypothetical rebellion allele, and since nearby nests are at risk of being raided by the same slavemaker nest, rebellion provides an opportunity for enslaved workers to increase the fitness of related nests.

The fact that we find enslaved workers and free-living individuals similarly related to surrounding colonies suggests the relative stability of local population structure, because the enslaved workers we genotyped were stolen as brood and are at least one year old. This also suggests, more generally, that slave raiding activity does not greatly disturb host population structure. For example, if hosts would move away after slavemaker contact, we would expect lower relatedness of enslaved workers to surrounding colonies, as compared to free-living nests and their neighbours. We therefore conclude that the regular raiding events do not disturb local population structure on the relevant time scale, i.e., during the lifetime of enslaved workers. This result is somewhat unexpected because *Temnothorax* ants inhabit ephemeral nest sites such as acorns, and they frequently relocate their nest during the summer. Moreover, new colonies are founded by young queens, which mated in large mating swarms (Howard and Kennedy, 2007). Both processes could erode population structure on the studied spatial scale.

Next, we determined that polydomy, the simultaneous occupation of more than one nest site and frequent exchange of workers, brood and queens (Debout et al., 2007), is most likely the main source for the observed structure. *Temnothorax* colonies are often polydomous (Alloway et al., 1982; Herbers and Tucker, 1986), which provide these ants with numerous benefits, including the optimal exploitation of patchy resources, increased survival chances of the colony in case of predation or stochastic events (Alloway et al., 1982) and perhaps increased productivity (Kramer et al., 2013). Our analysis indicates that polydomy is sufficient to explain the observed small-scale population structure, and can explain why enslaved workers are related to nearby colonies.

Polydomy leads to small-scale population structure and may therefore have allowed the evolution of slave-rebellion behaviour. In addition, polydomy may also function as a bet-hedging strategy, which allows a part of the colony to survive if the other part is attacked and destroyed in a raid. Polydomy would also allow host workers that escaped during a raiding attack to seek refuge in the secure nest. Polydomy may therefore serve as a defence trait against slavemaking ants, and if so, we would expect higher rates of polydomy in parasitised areas, if sufficient suitable nest sites were available. In support of this prediction, slavemaker presence led to a reduction in the number of workers and queens per nest (indicative of increased polydomy) in New York in a field manipulation experiment (Foitzik et al., 2009).

Thompson's work on coevolution (Thompson, 1994) has been very influential and established that population structure can shape host-parasite interactions, since spatial structure determines which hosts interact with which parasites. However, most studies on population structure and host-parasite interactions focus on a much larger spatial scale than the current study (Dybdahl and Lively, 1996; Sundström et al., 2005; Trontti et al., 2005b; Keeney et al., 2009). Previous studies by ourselves have also focused on larger scales (e.g., comparing the NY population with the WV population (Foitzik and Herbers, 2001a), and showing that *T. longispinosus* is structured on a large scale in North America (Pennings et al., 2011)). In this study, we looked at the smallest scale possible: the scale of individual nests. We observe strong structure at the smallest scale, finding that nests are closely related to neighbouring nests within a distance of 1 to 2 meters. After about 2 meters, the effect disappears and the population appears to be panmictic. This offers a potential explanation why we do not detect a significant overall population structure in WV using a mantel test. Our current and previous results can be explained by a combination of mating flights (causing panmixia at the level of the forest, but not at the level of North America) and polydomy (causing small scale population structure). Such variation of population structure on different scales can profoundly influence the outcome of the host-parasite co-evolutionary dynamic, and our studies show that the choice of scale and analysing method for a population genetic study will greatly influence the results.

Recent work suggests that in addition to the inherent influence of the co-evolutionary dynamic, two important parameters shape parasite virulence and host resistance: Host population structure and the spatial range of host-parasite interaction (Kamo et al., 2007; Lion and Baalen, 2008; Best et al., 2011). These studies concluded that viscous host populations in combination with local host-parasite interaction could select for lower virulence (prudence), while well-mixed populations and global host-parasite interactions should result in increased parasite virulence. The same rules could apply in our host-parasite system. As found in micro-parasites, the virulence of social parasites varies between species (e.g., Mori et al., 2001; Brandt et al., 2005; Johnson and Herbers, 2006), even between the closely related slavemakers *P. americanus* (the slavemaker in the current study) and *T. duloticus*, which share a similar ecology and an overlapping set of host species (Alloway, 1979; Mori et al., 2001; Johnson and Herbers, 2006). Both of these parasites harm their host species, but *T. duloticus* is extremely destructive (virulent), while the impact of *P. americanus* appears to be more moderate (Alloway, 1979; Hare and Alloway, 2001; Beibl et al., 2005). Our current study concludes that *P. americanus* is characterised by a limited raiding range, and its main host is characterised by a strong and stable population structure – exactly the conditions that are expected to lead to a prudent parasite. It would be of interest to see how *T. duloticus* compares to *P. americanus* in these respects. Another reason for the observed prudence in *P. americanus* could be slave rebellion itself. A key difference between the two slavemaker species is the number of slavemaker workers per nest. *T. duloticus* has, on average, five times as many workers as *P. americanus* (Beckers et al., 1989; Foitzik and Herbers, 2001a). We know that nest size is positively correlated to virulence in *P. americanus,* and expect that this may also be the case for *T. duloticus.* The low number of *P. americanus* workers per nest is at least partly caused by slave rebellion (up to 70% mortality rate, (Pamminger et al., 2013) and this aspect of host defence could explain the observed low virulence of *P. americanus*. It remains to be investigated whether slave rebellion occurs in *T. duloticus* or the recently described congener *T. pilagens* (Seifert *et al.*, 2014).

**Conclusions**

The analysis of our simple mathematical model led to the prediction that the local population structure was an important condition for the potential evolution of the slave rebellion trait. In a population genetic study, we then found a stable host population structure on a fine scale, resulting in an elevated relatedness of enslaved host workers to surrounding host colonies. This genetic structure is a necessary (but not sufficient) condition for indirect selection to act on the slave rebellion trait and presents a possible escape from an evolutionary dead end. Our study highlights the importance of spatial scale when analysing host-parasite interactions and provides empirical support for the potential role of population structure and life history traits in two co-evolving species.

**Acknowledgments**

This work was funded by the Deutsche Forschungsgemeinschaft Research Unit 1078 grants TP Fo 298 / 9-1, 2 and ME 3134/5-2 and 3134/6-1 (DFG SPP 1590), and by the E.N. Huyck Preserve. We thank Stephan Suette and Andreas Modlmeier for the help in the field, and Marion Kever, Yvonne Cämmerer and Bettina Rinjes for the help with DNA extraction and genotyping. In addition, we like to thank the three reviewers for their time invested in the improvement of this manuscript.

Table 1: Parameters of a simple mathematical model to discuss evolutionary stability of the absence of slave rebellion and the default settings used for the calculations (Fig.4).

| | | |
|---|---|---|
| *s* | 0.3 | Probability that a slavemaker nest arrives (if there are none initially) at the neighbourhood in one year |
| *d* | 0.1 | Probability that a rebel-free slavemaker dies at the end of the season. |
| *z* | 0.1 | Probability of a raided host nest to survive all raids of the season. |
| *r* | 0.8 | Probability that a slavemaker nest that has raided a host nest containing rebels dies at the end of the season. |
| *p* | 0.7 | Probability to be raided by a slavemaker nest, if it is in the neighbourhood. |
| *n* | 0.5 | Average number of other rebel hosts nests in a neighbourhood of a focal rebel host nest. |
| *u* | 0.9 | Probability that a host nest does not die for other reasons than raids. |
| *c* | – | Cost of rebellion. That is, the productivity of a rebel nest is lower than the productivity of a non-rebel host nest by a factor of *c*. |

**Figures**

Figure 1. Exemplary distribution of plots within a transect in WV. In the lower right corner plot R is enlarged to demonstrate the distribution of colonies within a plot, and to show the distribution of alleles of one of the six microsatellite loci (GT1) among colonies. Each colony is represented by a pie-diagram with the frequencies of different GT1 alleles amongst the genotyped individuals of the colony. R3 is a slavemaker nest and shares most of its alleles with the free, queenright, nest R7. R13 and R15 are free-living host colonies in close proximity and appear to be related. Colonies of other species are not shown. All units are in meters (m).

Figure 2. Distribution of relatedness values a) top chart: within nests $D_{within}$ (green), b) middle chart: between independent nests $D_{indep}$ (black), c) bottom chart: fitted mixture (blue) of $D_{within}$ (green) and the $D_{indep}$ (black) for nest pairs at less than 1m distance. The left column shows data from New York (NY), the right column data from West-Virginia (WV). The threshold above which nest pairs are more likely to be polydomous is indicated with a blue dashed line.

Figure 3. The relatedness between nests of free-living host colonies in New York (upper chart) and West Virginia (lower chart) indicating elevated relatedness when colonies are close to each other. The colonies above the dashed blue line are likely to be polydomous, exhibiting high degrees of relatedness in close proximity to each other.

Figure 4. The effect of model parameters on the fitness of the host nests containing rebels, where fitness is relative to non-rebel host nests. Cost of rebellion was set to 0.05 (dotted line), 0.1 (dashed line) and 0.2 (solid line). For each plot one model parameter was varied, and all other parameters were set to the default parameter settings indicated by the vertical dashed line in the corresponding plot (see also Table 1).

Figure 5. The average relatedness of enslaved workers to host colonies within raiding range (red) in contrast to host colonies further away (black). Data from New York are presented on the left, and data from West Virginia on the right. Mean relatedness values are presented and error bars indicate standard error.

**Figure 1:**

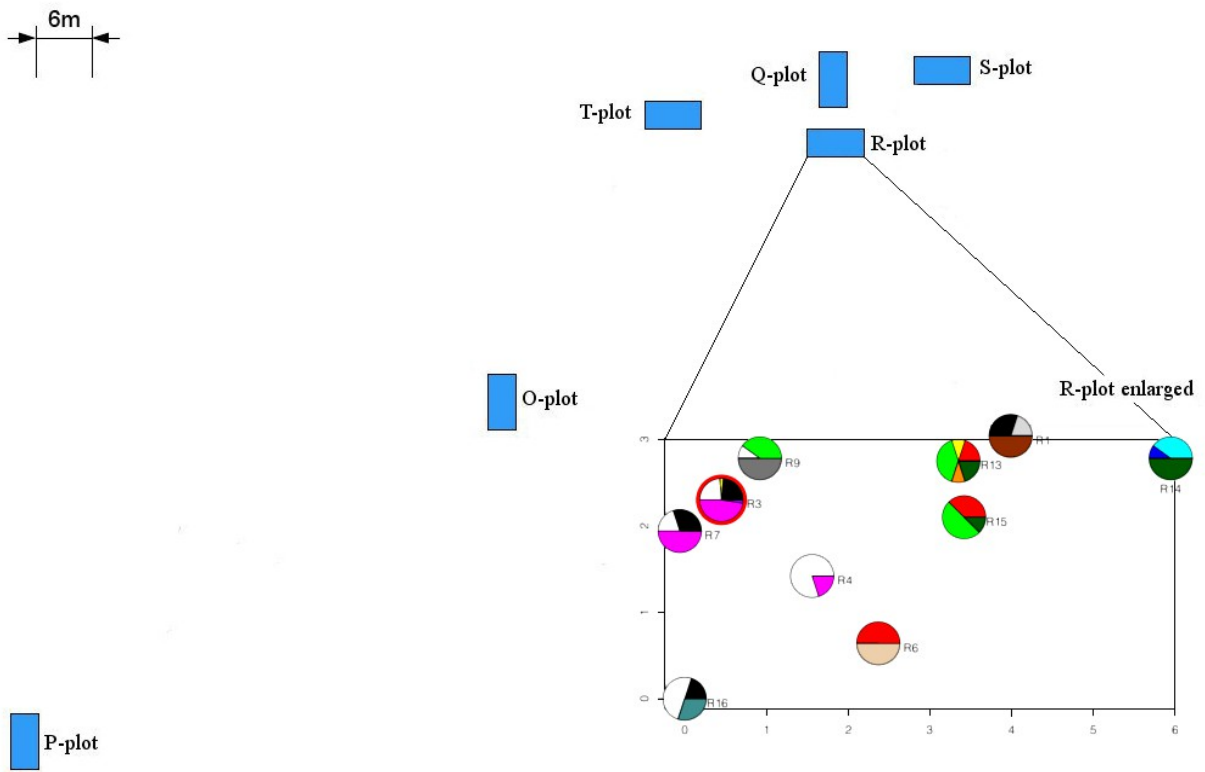

**Figure 2:**

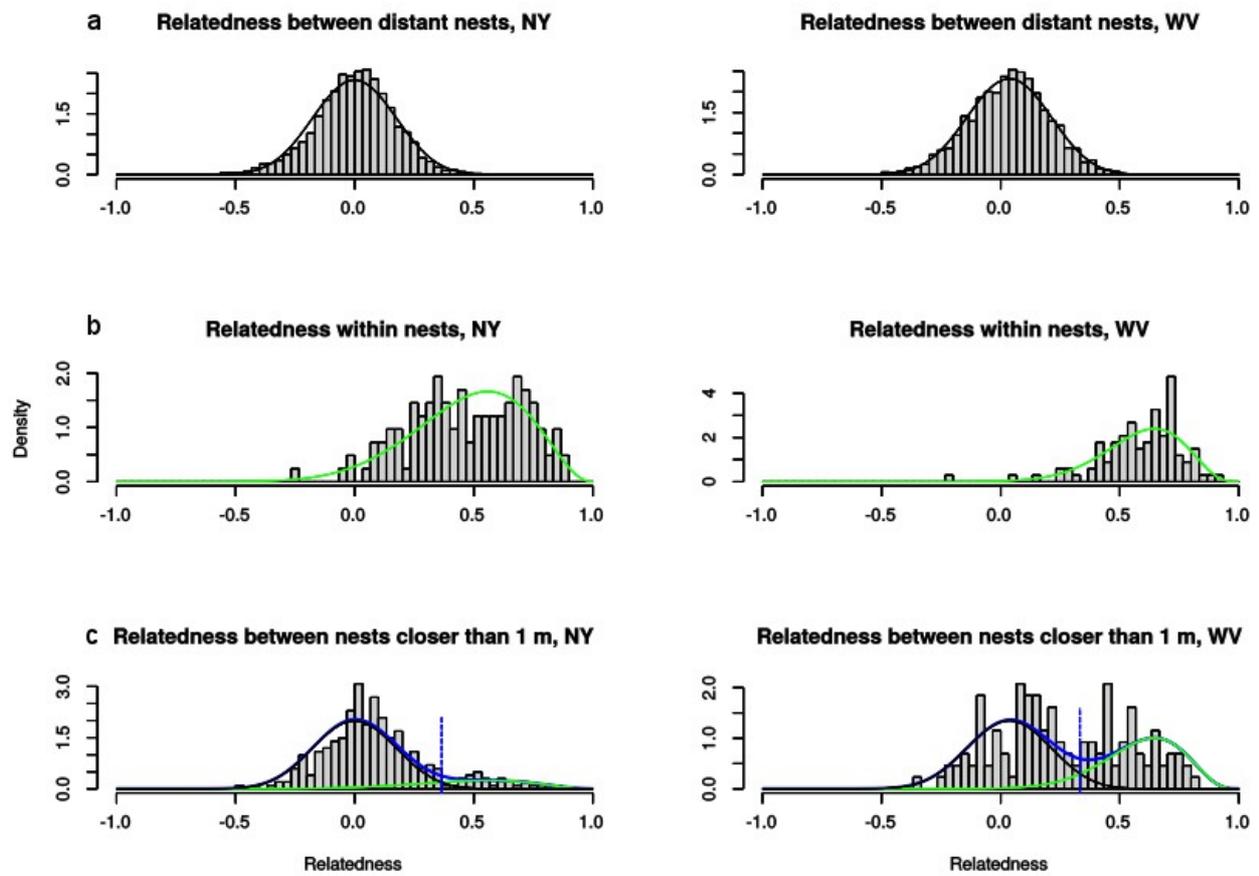

**Figure 3:**

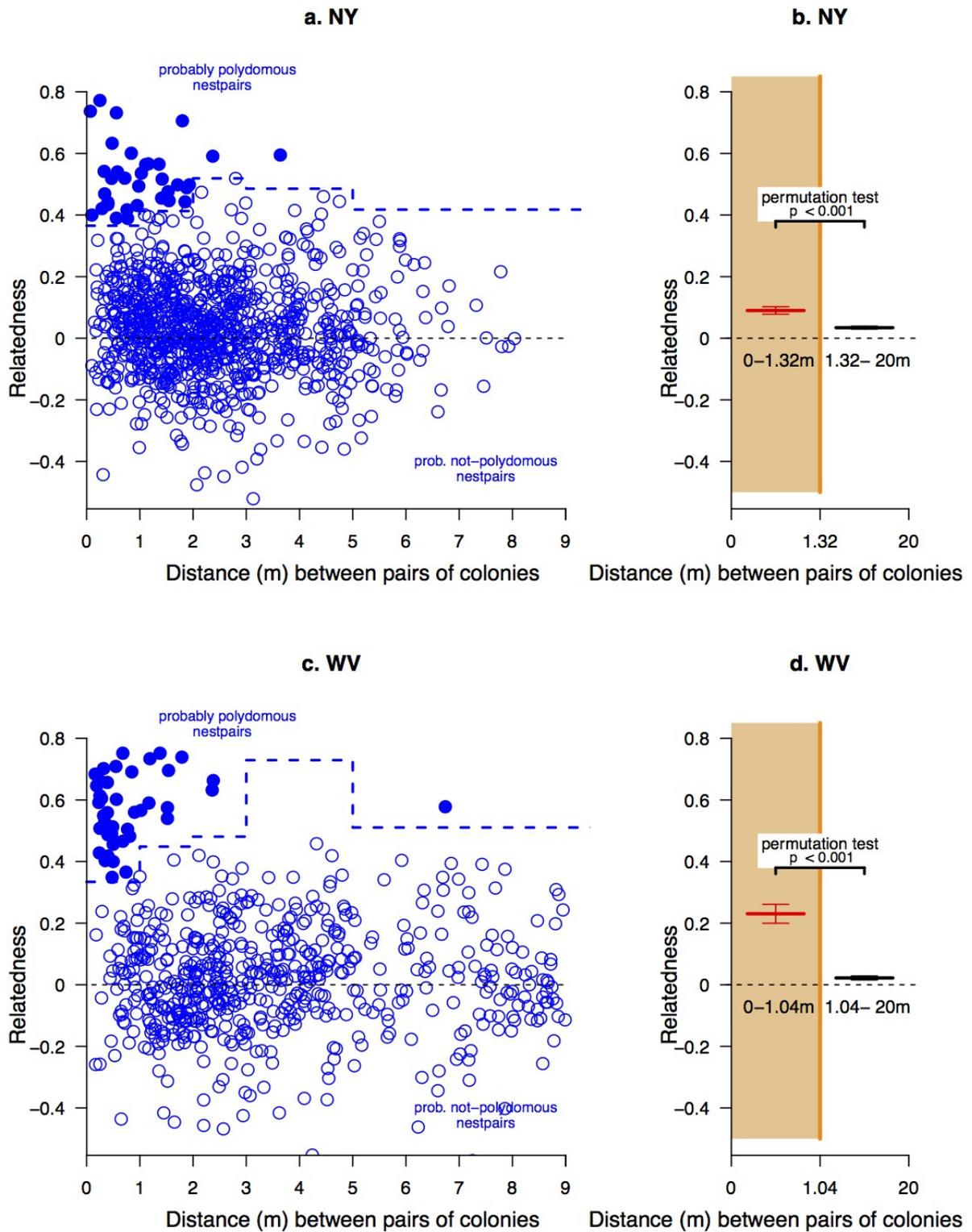

**Figure 4:**

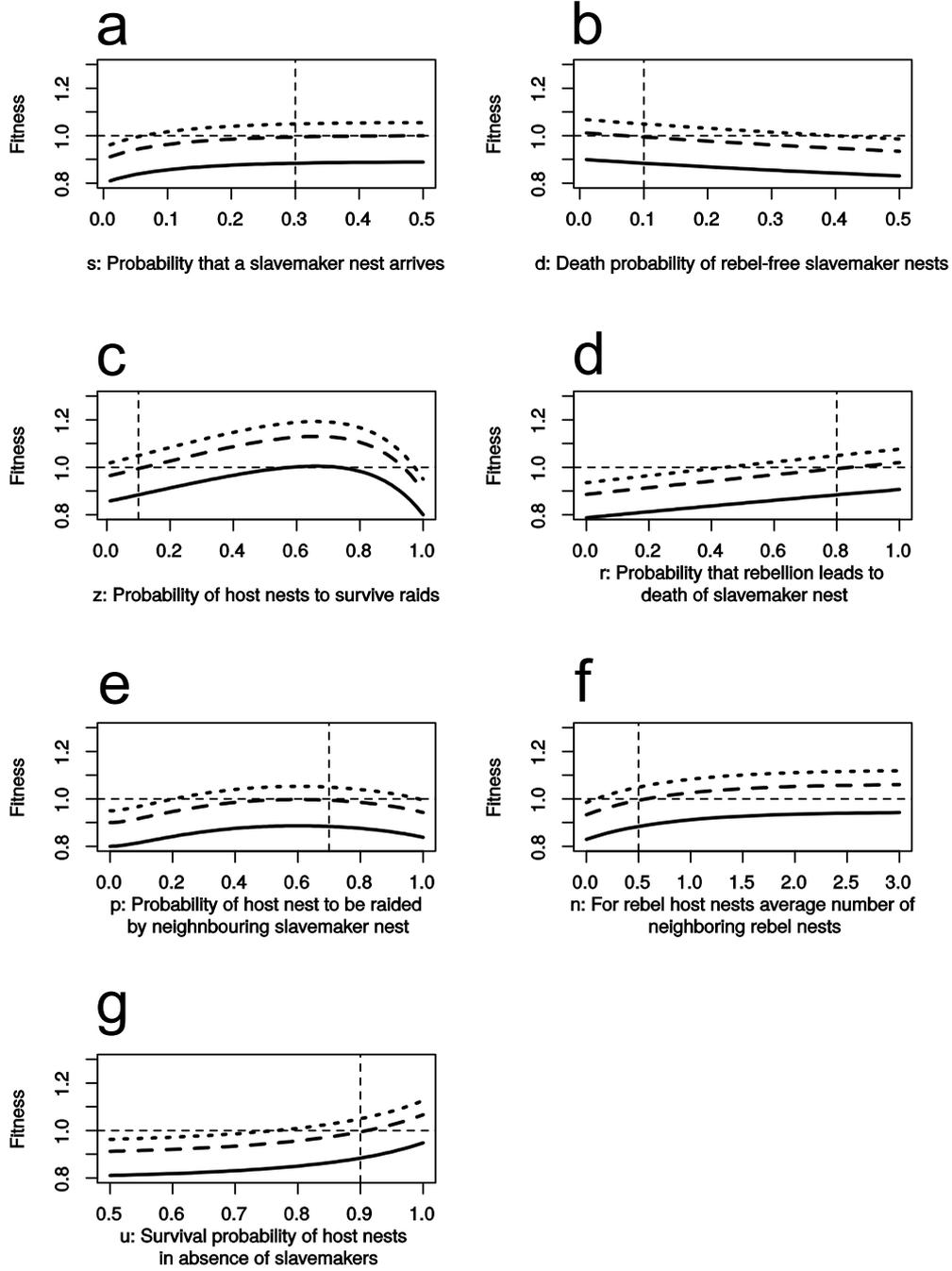

**Figure 5:**

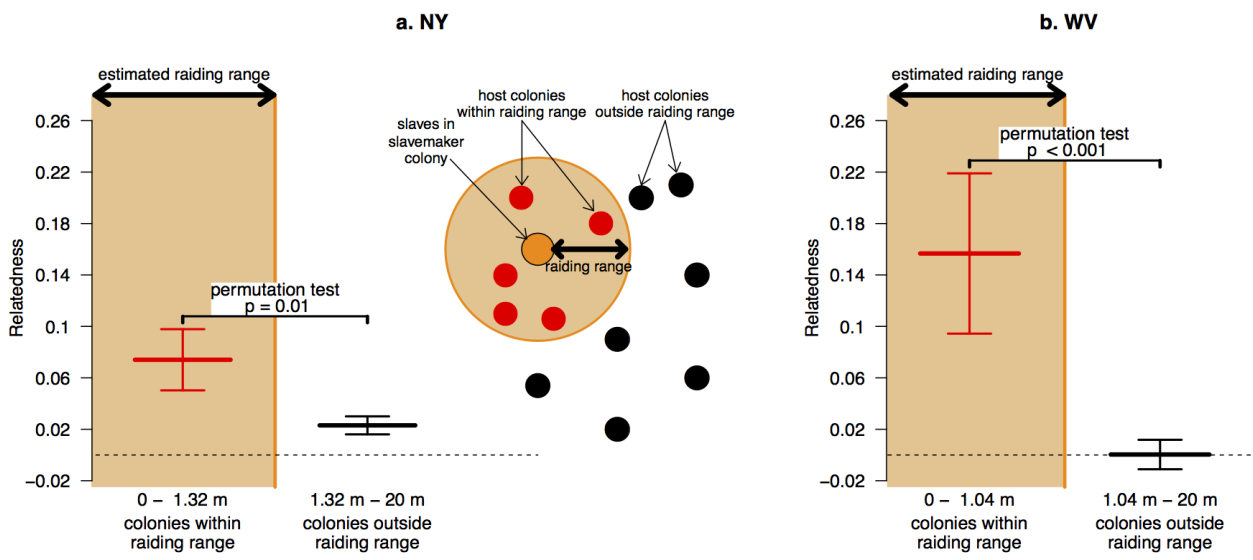